\documentclass[aps,pre,twocolumn,groupedaddress,showpacs,floatfix]{revtex4}
\usepackage{dcolumn}
\usepackage{amsmath}
\usepackage{amssymb}
\usepackage{graphicx}
\usepackage{bm}
\usepackage[T1]{fontenc}
\usepackage{color}
\usepackage{url}
\usepackage[bf]{subfigure}
\usepackage{rotating}
\usepackage{scalefnt}

\begin{document}

\title{On degree-degree correlations in multilayer networks}

\author{Guilherme Ferraz de Arruda}
\affiliation{Departamento de Matem\'{a}tica Aplicada e Estat\'{i}stica, Instituto de Ci\^{e}ncias Matem\'{a}ticas e de Computa\c{c}\~{a}o,
Universidade de S\~{a}o Paulo - Campus de S\~{a}o Carlos, Caixa Postal 668,
13560-970 S\~{a}o Carlos, SP, Brazil.}

\author{Emanuele Cozzo}
\affiliation{Institute for Biocomputation and Physics of Complex Systems (BIFI) \& Department of Theoretical Physics, University of Zaragoza, 50018 Zaragoza, Spain}

\author{Yamir Moreno}
\affiliation{Institute for Biocomputation and Physics of Complex Systems (BIFI) \& Department of Theoretical Physics, University of Zaragoza, 50018 Zaragoza, Spain}
\affiliation{Complex Networks and Systems Lagrange Lab, Institute for Scientific Interchange, Turin, Italy}

\author{Francisco A. Rodrigues}
\email{francisco@icmc.usp.br}
\affiliation{Departamento de Matem\'{a}tica Aplicada e Estat\'{i}stica, Instituto de Ci\^{e}ncias Matem\'{a}ticas e de Computa\c{c}\~{a}o,
Universidade de S\~{a}o Paulo - Campus de S\~{a}o Carlos, Caixa Postal 668,
13560-970 S\~{a}o Carlos, SP, Brazil.}

\begin{abstract} 
We propose a generalization of the concept of assortativity based on the tensorial representation of multilayer networks, covering the definitions given in terms of Pearson and Spearman coefficients. Our approach can also be applied to weighted networks and provides information about correlations considering pairs of layers. By analyzing the multilayer representation of the airport transportation network, we show that contrasting results are obtained when the layers are analyzed independently or as an interconnected system. Finally, we study the impact of the level of assortativity and heterogeneity between layers on the spreading of diseases. Our results highlight the need of studying degree-degree correlations on multilayer systems, instead of on aggregated networks.
\end{abstract}


\maketitle

\section{Introduction}

The use of network science to study the structure and dynamics of complex systems has proved to be a successful approach to understand the organization and function of several natural and artificial systems~\cite{Boccaletti06:PR, Costa07:AP, Newman010:book, Costa011:AP}. The traditional framework used up to a few years ago represents the structure of complex systems as single-layer (also referred to as monoplex) networks, in which only one type of connection is accounted for. However, this approach is limited because most natural and artificial systems such as the brain, our society or modern transportation networks~\cite{Kivela2014, BoccalettiPR2014}, are made up by different constituents and/or different types of interaction. Indeed, their structure is organized in layers. For instance, in social networks individuals can be connected according to different social ties, such as friendship or family relationship~(e.g.~\cite{Verbrugge79}). In transportation networks, routes of a single airline can be represented as a network, whose vertices (destinations) can be mapped into networks of several companies~\cite{Cardillo2013}. Gene co-expression networks consist of layers, each one representing a different signaling pathway or expression channel~\cite{Li011:PLoS}. Therefore, mapping out the structure of these and similar systems as a monoplex network could lead to miss relevant information that could not be captured if the single layers are analyzed separately nor if all layers are collapsed altogether in an aggregated graph. Additionally, note that in most of these interconnected systems, the information travels not only among vertices of the same layer, but also between pairs of layers.

Recent advances in modeling the aforementioned systems include new mathematical formulations~\cite{PhysRevX.3.041022}, the generalization of different metrics ~\cite{BoccalettiPR2014, PhysRevX.3.041022, DeDomenicoPNAS2014, 2013arXiv1307.6780C} and the impact of the multilayer structure on several dynamical processes \cite{Cozzo:2013, DeDomenicoPNAS2014, BuldyrevN2010, GomezPRL2013, GranellPRL2013}.
Although clustering~\cite{2013arXiv1307.6780C}, centrality~\cite{DeDomenicoPNAS2014, 1311.2906v1} and spectral properties~\cite{Cozzo:2013, DeDomenicoPNAS2014, SolePRE2013} of multilayer networks have been addressed, a measure to quantify degree-degree correlations in multilayers is still lacking. Degree-degree correlations is a fundamental property of single-layer networks, impacting the spreading of diseases, synchronization phenomena and systems' resilience~\cite{Barrat08:book, Newman010:book}. Additionally, it has been reported that different correlations arise in different kinds of networks: social networks are in general assortative, meaning that highly connected nodes tend to link with each other, whereas technological and biological systems have disassortative structures, in which high degree nodes are likely attached to low degree nodes~\cite{newman02assortative}.

For networks made up of more than one layer, only recently, Nicosia and Latora~\cite{2014arXiv1403.1546N} considered the correlation between the degrees in two different layers. However, their methodology is only for node-aligned multiplex networks, which are special cases of multilayer networks (see~\cite{Kivela2014}). In fact, multiplex networks are made up of $N$ nodes that can be in one or more interacting layers. The links in each layer represent a given mode of interaction between the set of nodes belonging to that layer, whereas links connecting different layers stand for the different modes of interaction between objects involved in~\cite{Kivela2014}. 

In this paper we study degree-degree correlations in multilayer systems and propose a way to generalize previous assortativity metrics by considering the tensorial formulation introduced in~\cite{PhysRevX.3.041022}. Our approach also covers a weighted version of assortativity~\cite{Leung2007} and the case in which the assortativity is given by the Spearman correlation coefficient, generalizing the definition in~\cite{PhysRevE.87.022801}. The study of a real dataset corresponding to the airport transportation network shows a contrasting behavior between the analyses of each layers independently and altogether, which reinforces the need for such a generalization of the assortativity measure. Finally, we study the influence of degree-degree correlations on epidemic spreading in multilayer networks. We verify that the impact of the disease depends on degree-degree correlations and also on the level of heterogeneity between the layers.

\section{Assortativity in multilayer networks.}

Tensors are suitable for representation of multilayer networks. As showed in~\cite{PhysRevX.3.041022}, tensors allow us to consider a branch of new relationships between nodes and layers, by encoding a multilayer network as a forth order mixed tensor, $M_{\beta \tilde{\gamma}}^{\alpha \tilde{\delta}}$, i.e. 2-covariant and 2-contravariant basis, in the Euclidian space. Such representation is convenient for many operations, as discussed in~\cite{PhysRevX.3.041022}. We use the definition of the interlayer adjacency tensor $C_\beta^\alpha(\tilde{h} \tilde{r})$ that is a second order tensor which has the information of the relationships between nodes in layers $\tilde{h}$ and $\tilde{r}$. Note that $C_\beta^\alpha(\tilde{r} \tilde{r})$ is the adjacency matrix for the layer $\tilde{r}$ and belongs to $\mathbb{R}^{N \times N}$ space. Then, the multilayer adjacency tensor is expressed as the summation over all layers $L$ of the tensorial product of the adjacency tensors, $C_\beta^\alpha(\tilde{h} \tilde{r})$, and the canonical Euclidean basis. Mathematically, 
\begin{equation}
 M_{\beta \tilde{\delta}}^{\alpha \tilde{\gamma}} = \sum_{\tilde{h}, \tilde{r}}^{L} C_\beta^\alpha(\tilde{h} \tilde{r}) E_{\tilde{\gamma}}^{\tilde{\delta}}(\tilde{h} \tilde{r}).
\end{equation}
which belongs to $\mathbb{R}^{N \times N \times L \times L}$ space. 

Following Einstein's summation convention, the assortativity coefficient can be written as
\begin{widetext}
\begin{equation}
  \rho(\mathcal{W}_\beta^\alpha) = \frac{\mathcal{M}^{-1} \mathcal{W}_\beta^\alpha Q^\beta Q_\alpha -  \left[ 1/2 \mathcal{M}^{-1} \left( \mathcal{W}_\beta^\alpha Q_\alpha u^\beta + \mathcal{W}_\beta^\alpha Q^\beta u_\alpha \right) \right]^2}{\mathcal{M}^{-1} \left( \mathcal{W}_\beta^\alpha (Q_\alpha)^2 u^\beta + \mathcal{W}_\beta^\alpha (Q^\beta)^2 u_\alpha \right) - \left[1/2 \mathcal{M}^{-1} \left( \mathcal{W}_\beta^\alpha Q_\alpha u^\beta + \mathcal{W}_\beta^\alpha Q^\beta u_\alpha \right) \right]^2}
 \label{eq:multiplex}
\end{equation}
\end{widetext}
where $u$ is the 1-tensor, which is a tensor of rank 1 and has all elements equal to 1, $\mathcal{W}_\beta^\alpha$ is a second order tensor that summarizes the information that is being extracted and $\mathcal{M} = \mathcal{W}_\beta^\alpha U^{\beta}_{\alpha}$ is a normalization constant. 

Let us explain in more details all terms appearing in the expression of $  \rho(\mathcal{W}_\beta^\alpha)$. First, we define
\begin{equation}
 Q^\alpha = \mathcal{W}_\beta^\alpha u^\beta,
\label{eq:Qa}
\end{equation} 
which is a 1-contravariant tensor and
\begin{equation}
 Q_\beta = \mathcal{W}_\beta^\alpha u_\alpha
\label{eq:Qb}
\end{equation}
which is a 1-covariant tensor. Moreover, the indices are related to the direction of the relationships between nodes. Such a choice ensures a more general expression, capturing degree correlations on non-symmetric tensors and, consequently, in directed and weighted networks.

Due to the multiplex nature of such systems we obtain different types of correlations, which can be uncovered by operating on the adjacency tensor. First of all, it is possible to extract a single layer by the operation called \emph{single layer extraction}~\cite{PhysRevX.3.041022}. In this case, the adjacency tensor is defined as
\begin{equation}
\mathcal{W}_\beta^\alpha = C_\beta^\alpha(\tilde{r} \tilde{r}) = M_{\beta \tilde{\delta}}^{\alpha \tilde{\gamma}} E_{\tilde{\gamma}}^{\tilde{\delta}}(\tilde{r} \tilde{r}),
\label{eq:layer_ex}
\end{equation}
which is a simple projection on the canonical basis, $E_{\tilde{\gamma}}^{\tilde{\delta}}(\tilde{r} \tilde{r})$. It is noteworthy that the results obtained from this projection are the same as those obtained by considering the layer $\tilde{r}$ as a monoplex network and applying the traditional formulation of assortativity~\cite{newman02assortative}. On the other hand, to consider all layers altogether, we can use the \emph{projected network}, which is a weighted single-layer network. Formally it is given as
\begin{equation}
\mathcal{W}_\beta^\alpha = P_\beta^\alpha = M_{\beta \tilde{\gamma}}^{\alpha \tilde{\delta}} U_{\tilde{\delta}}^{\tilde{\gamma}}.
\label{eq:projection}
\end{equation}
Note that the projection presents self-edges and, as argued in~\cite{PhysRevX.3.041022}, it is different from a weighted monoplex network, since self-edges code for inter-layer couplings between different replica of the same object. Thus they have a different meaning with respect to other edges. A version of the projection without self-edges is called \emph{overlay network} and is given as the contraction over the layers~\cite{PhysRevX.3.041022}, i.e.,
\begin{equation}
\mathcal{W}_\beta^\alpha = O_\beta^\alpha = M_{\beta \tilde{\gamma}}^{\alpha \tilde{\gamma}}.
\end{equation}
Observe that the overlay network does not consider the contribution of the interlayer connections, whereas the projection does. As we will see later, comparisons between the assortativity of those two different representations of the system reveal the key role of such inter-links.  

In both cases, i.e., for the overlay and the projected networks, we extract degree-degree correlations. Nodes with similar degrees connected in the same or different layers contribute positively to the assortativity coefficient. On the other hand, the connections between hubs and low degree nodes in the same or different layers decrease the assortativity. Self-edges always increase the assortativity, which yields different values of assortativity for the overlay and the projected networks. This gives information on the nature of the coupling between different replicas of the same object among different layers.

In some applications, it is interesting to calculate a pair-wise correlation between a set of nodes, for instance, between couple of layers. Thus, we propose a new operation, that we call \emph{selection}, which is a projection over a selected set of layers:  
\begin{equation}
\mathcal{W}_\beta^\alpha(\mathcal{L}) = S_\beta^\alpha(\mathcal{L}) = M_{\beta \tilde{\gamma}}^{\alpha \tilde{\delta}} \Omega_{\tilde{\delta}}^{\tilde{\gamma}}(\mathcal{L}),
\label{eq:selection}
\end{equation}
where $\Omega_{\tilde{\delta}}^{\tilde{\gamma}}$ is a tensor used to select the set of layers we consider in the projection ($\mathcal{L}$). The components of the tensor are equal to unity when the layers $\tilde{\delta}$ and $\tilde{\gamma}$ are selected, and zero otherwise. Note that by selecting all layers together we recover the 1-tensor $U_{\tilde{\delta}}^{\tilde{\gamma}}$ and consequently Eq.~\ref{eq:projection}. Another special case is $\tilde{\delta} = \tilde{\gamma}$, which yields Eq.~\ref{eq:layer_ex}, or the layer extraction. The tensor can also be generalized to weight different layers. In this case, each element of $\Omega_{\tilde{\delta}}^{\tilde{\gamma}}$ contains the weight of the relationship between two layers $\tilde{\delta}$ and $\tilde{\gamma}$.
Such projection is similar to the covariance matrix in statistics, which generalizes the concept of variance. The covariance between two variables is quantified in each entry of the matrix and the main diagonal has the variance of each variable. Thus, we can define a matrix that generalizes the assortativity in a similar manner as the covariance matrix generalizes the concept of variance, i.e.
\begin{equation}
\mathbf{S}^{\tilde{\gamma}}_{\tilde{\delta}} = \rho \left( S^\alpha_\beta(\mathcal{L} = \{ \tilde{\gamma}, \tilde{\delta} \}) \right),
\label{eq:selection_matrix}
\end{equation}
which belongs to a $\mathbb{R}^{L \times L}$ space. We call $\mathbf{S}$ the \textit{P-assortativity matrix}.

Also in this case, a similar operation for the overlay network can be considered, yielding
\begin{equation}
\mathcal{W}_\beta^\alpha(\mathcal{L}) = Z_\beta^\alpha(\mathcal{L}) = \sum_{\tilde{h} \in \mathcal{L}}^L C_{\beta}^{\alpha} (\tilde{h} \tilde{h}),
\end{equation}
which can also be generalized in a similar way as Eqs.~\ref{eq:selection} and~\ref{eq:selection_matrix}, resulting in the matrix 
\begin{equation}
\mathbf{Z}^{\tilde{\gamma}}_{\tilde{\delta}} = \rho \left( Z^\alpha_\beta(\mathcal{L} = \{ \tilde{\gamma}, \tilde{\delta} \}) \right).
\end{equation}\label{eq:Z}
We call $\mathbf{Z}$ the \textit{O-assortativity matrix}. A similar interlayer correlation was also proposed in~\cite{2014arXiv1403.1546N}, where the authors suggested measuring the degree correlation between two different layers of the replica of the same object (or node). Furthermore, they proposed three different ways: the Pearson correlation coefficient, Spearman rank correlation and the Kendall's $\tau$ index. However, it is worth pointing out that such an approach does not consider the intra-layer relationship because it is only for node-aligned multiplex networks \cite{Kivela2014}. Here, we generalize such a measure in terms of tensorial notation.

Up to now we have considered nodes, but if we extract the \emph{network of layers}~\cite{PhysRevX.3.041022}, the correlation between different layers can also be evaluated. We use
\begin{equation}
\mathcal{W}_{\tilde{\delta}}^{\tilde{\gamma}} = \Psi_{\tilde{\delta}}^{\tilde{\gamma}} = M_{\beta \tilde{\delta}}^{\alpha \tilde{\gamma}} U^{\beta}_{\alpha}
\end{equation}
where $U^{\beta}_{\alpha}$ is the second-order tensor whose components are all equal to one. It is important to stress that the components of this adjacency tensor are not binary, but weighted by the number of edges inter each layer. Moreover, also in this case, the resulting tensor presents self-edges that encode the information about the density of connections inside a single layer. Finally, we can consider only interlayer relationships over two different layers. Such information is extracted by projecting the adjacency tensor on the canonical base as
\begin{equation}
\mathcal{W}_\beta^\alpha = C_\beta^\alpha(\tilde{r} \tilde{h}) = M_{\beta \tilde{\gamma}}^{\alpha \tilde{\delta}} E_{\tilde{\delta}}^{\tilde{\gamma}}(\tilde{r} \tilde{h}).
\label{eq:2l}
\end{equation}
Note that this is only applicable to multilayer networks and does not make sense in multiplex networks, since in the latter case the coupling is diagonal.

The assortativity coefficient can also be defined in terms of the Spearman rank correlation~\cite{PhysRevE.87.022801}, since the traditional definition of this coefficient based on the Pearson correlation~\cite{newman02assortative} can lead to incomplete results, as discussed in ~\cite{PhysRevE.87.022801}. The generalization of assortativity coefficient proposed here allows to consider the Spearman rank correlation coefficient by changing Eqs.~\ref{eq:Qa} and~\ref{eq:Qb}. Specifically, instead of considering the values of $Q^\alpha$ and $Q_\beta$, one substitutes them by their respective ranks. Such transformation is performed by using
\begin{equation}
Q^\alpha = \text{rank}(\mathcal{W}_\beta^\alpha u^\beta)
\end{equation}
and
\begin{equation}
Q_\beta = \text{rank}(\mathcal{W}_\beta^\alpha u_\alpha),
\end{equation}
where $\text{rank}(X_i)$ is the rank of the tensor $X_i$.

We henceforth denote by $\rho^P(\mathcal{W}_\beta^\alpha)$ and $\rho^S(\mathcal{W}_\beta^\alpha)$ the Pearson and Spearman correlation coefficients, respectively. Furthermore, we adopt $(\mathbf{S}^P)^{\tilde{\gamma}}_{\tilde{\delta}}$ and $(\mathbf{S}^S)^{\tilde{\gamma}}_{\tilde{\delta}}$ for the pair-wise correlation matrices using the Pearson and Spearman correlation coefficients, respectively. The same notation can be used for the matrices $(\mathbf{Z}^P)^{\tilde{\gamma}}_{\tilde{\delta}}$ and $(\mathbf{Z}^S)^{\tilde{\gamma}}_{\tilde{\delta}}$. Monoplex assortativity, i.e. assortativity in single-layer networks~\cite{newman02assortative}, is recovered by considering the adjacency matrix, $\mathcal{W}_\beta^\alpha = A_\beta^\alpha$, and consequently $Q^\alpha$ and $Q_\beta$ are analogous to in-degree and out-degree, respectively. Note that $Q^\alpha = Q_\beta$ for undirected networks. Moreover, $\mathcal{M}$ is equal to twice the number of edges, recovering the equation introduced in~\cite{newman02assortative}, which also captures correlations of weighted networks, as exposed in~\cite{Leung2007}.

\section{Application to real data}

\begin{figure*}[t]
\begin{center}
\includegraphics[width=2.1\columnwidth]{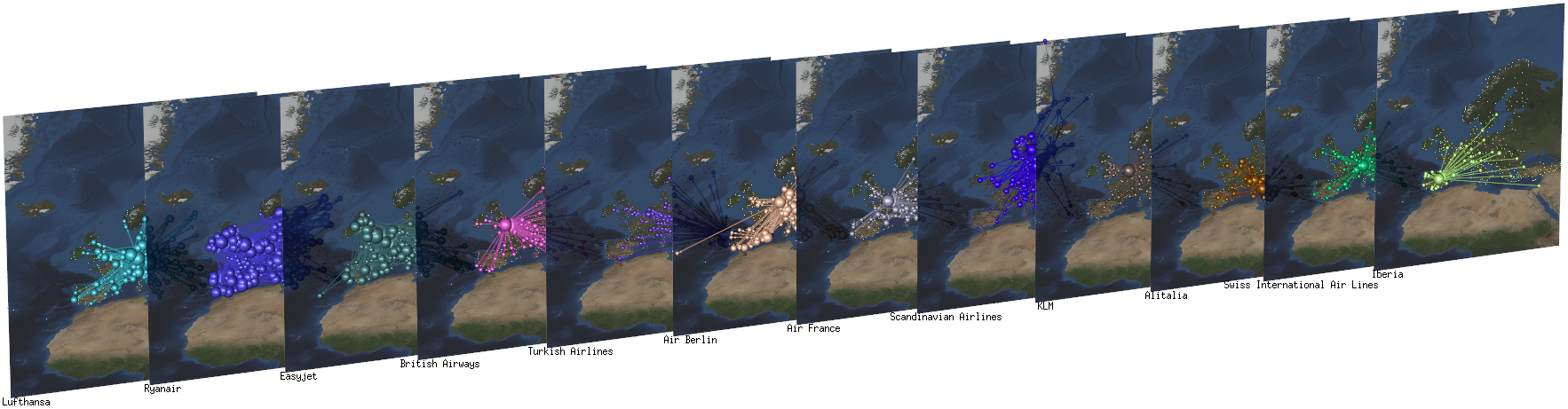}
\caption{Example of an airport transportation multilayer network. Each layer represents an airline, in which each node represents an airport and the edges are flights between two airports. This visualization was generated using MuxViz~\cite{DeDomenicoJCN2014}.}
\label{Fig:Airport_multilayer_network}
\end{center}
\end{figure*}

We analyze the airport transportation network \cite{Colizza:physics0602134}, whose multilayer representation was studied in~\cite{Cardillo2013}. The network comprises 450 airports and 37 companies, which are mapped as nodes and layers, respectively. More specifically, in each layer, the edges represent the directed flights operated by a given company and nodes, airports.  Figure~\ref{Fig:Airport_multilayer_network} shows a representation of 12 layers of such multilayer network. The inter-layer connections link the airports shared by pairs of different companies. 
This approach gives us a multilayer network that is not a node-aligned multiplex network, since the latter considers a diagonal coupling between all nodes in all layers. Note that the way proposed in~\cite{Cardillo2013} to create the aggregated monoplex network is the union of all layers considering multiple edges as single ones. This is in contrast to our approach, because we consider the projections and overlay networks as weighted networks, thus retaining the information of the number of different connections between the same pair of airports. 

Previous studies ~\cite{Colizza:physics0602134, Cardillo2013} showed that the airport transportation network presents the rich-club effect, which refers to the tendency of highly central nodes to be connected among themselves. This is also captured by the assortativity as shown in Table~\ref{tab:Global}, where we verify that the projected network has positive assortativity coefficients, agreeing with previous analyses. However note that the projection has a positive value of the assortativity, whereas the overlay has a negative one. Thus, the assortativity of the projection indicates that many companies share hubs airport, not that hubs connect between them. This apparent contradiction results from the fact that the rich-club effect is masked out in the overlay setup by the large number of peripheral nodes connecting to hubs. 

The analysis of each layer separately shows a different result, where most of the layers are disassortative. The only exception is the Netjet layer, which presents a positive coefficient for the rank correlation. Usually the companies focus their activities in one city or country, for example, Lufthansa in Germany or Air France in France, and have flights to other airports where their activity is lower. This leads to the disassortative behavior of each layer. Additionally, the disassortative correlations found in single layers is more pronounced than that of the overlay representation, which can be explained by noticing that hubs of a company are peripheral (or secondary) airports for other companies, but when the layers are collapsed they are also hubs in the overlay network and are connected.

\begin{figure*}[t]
\begin{center}
\subfigure[][]{\includegraphics[width=0.9\columnwidth]{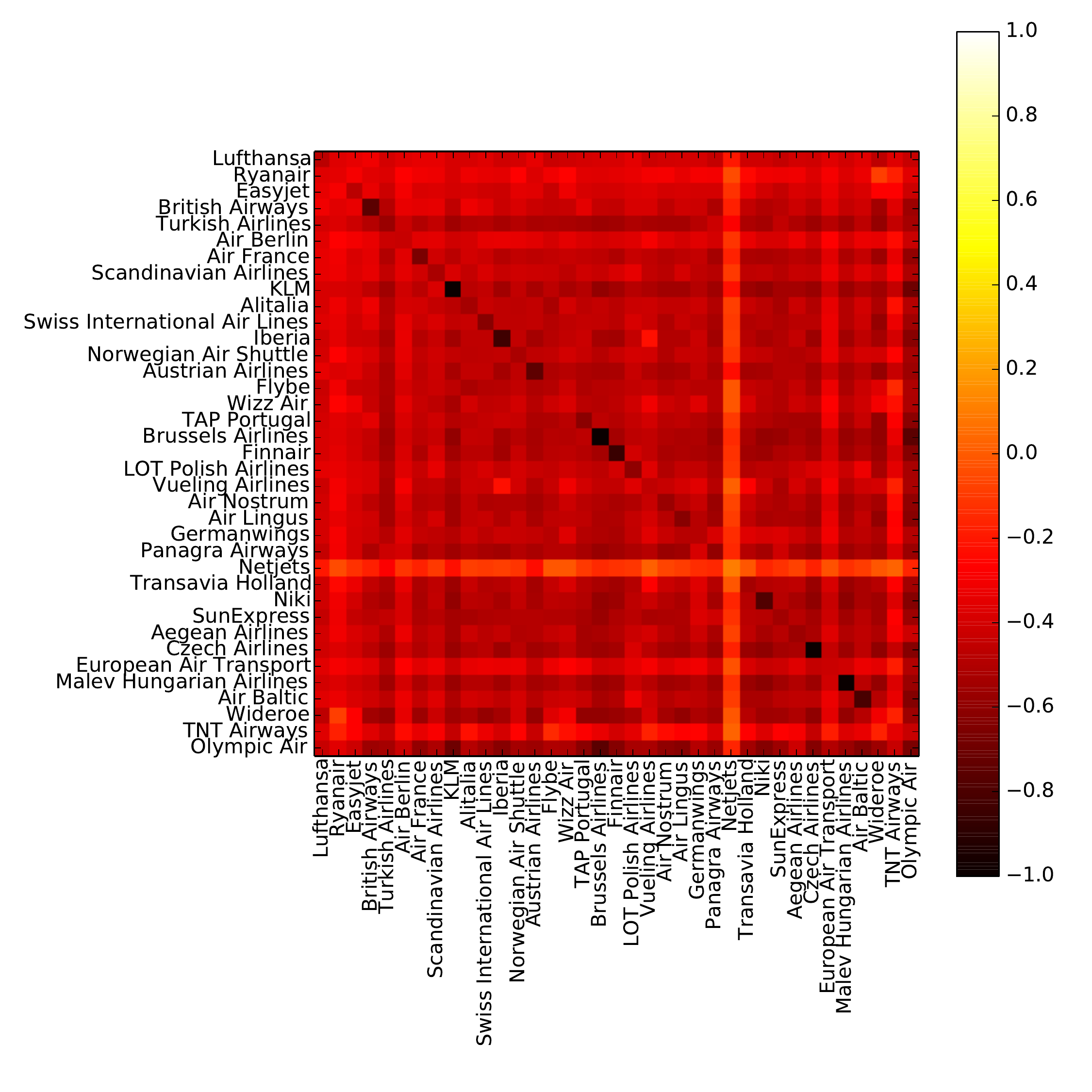}}
\subfigure[][]{\includegraphics[width=0.9\columnwidth]{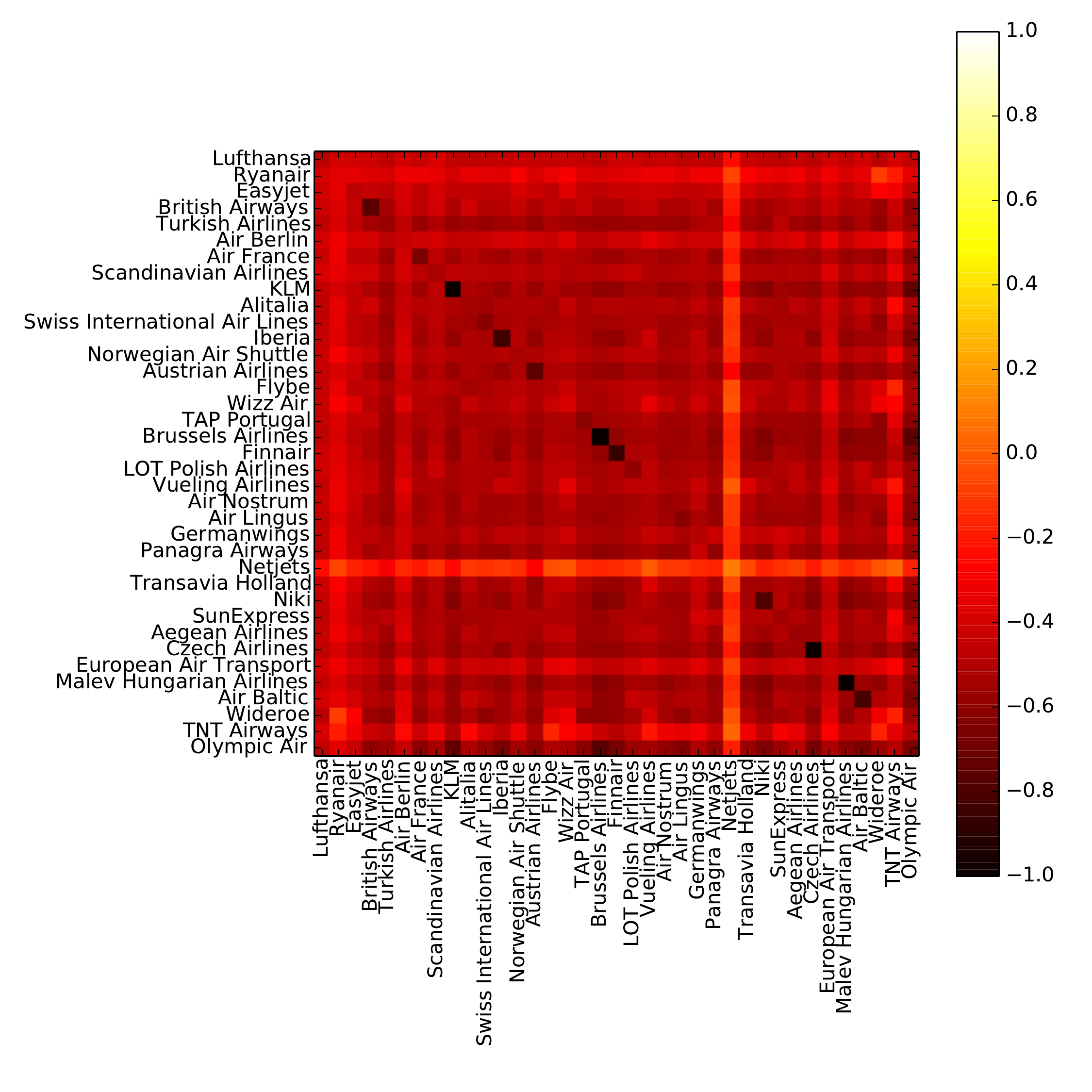}}
\caption{Pair-wise assortativity coefficient using Spearman rank correlation, $\rho^S(\mathbf{S}_\beta^\alpha)$  in (a) and $\rho^S(\mathbf{Z}_\beta^\alpha)$  in (b). Observe that the main diagonal presents the same coefficient considering the layer extraction operation, $\rho^S(C_\beta^\alpha(\tilde{r} \tilde{r}))$.}
\label{Fig:Airport_projection_rho}
\end{center}
\end{figure*}

Figure~\ref{Fig:Airport_projection_rho} shows the pair-wise correlation between layers. Interestingly, the latter is disassortative, in contrast to the results obtained for the projected network, but of the same sign as those computed for the overlay representation (see Table~\ref{tab:Global}). Furthermore, our construction of the adjacency tensor leads to an assortative network of layers, suggesting that bigger companies tend to share similar airports. This analysis agrees with~\cite{Cardillo2013}, where the authors argued that the main airports are connected to each other via directed flights. In addition, considering the Pearson correlations, the \textit{O-assortativity} matrix presents lower values if compared to the \textit{P-assortativity} matrix due to the intra-layer contributions, as discussed before.

\begin{table}[t] 
\begin{center}
\scalefont{0.85}
\caption{Structural properties of the airport transportation multilayer networks.}
\begin{tabular}{|l|c|c|c|c|c|}
\hline
Network & $N$ & $\mathcal{M}$ & $\langle Q^\alpha \rangle$ & $\rho^P(\mathcal{W}_\beta^\alpha)$ & $\rho^S(\mathcal{W}_\beta^\alpha)$ \\
\hline
Network of Layers $(\Psi_\beta^\alpha)$	& 37 & 30398.0 & 821.568 & 0.377 & 0.286 \\
Overlay $(O_\beta^\alpha)$	 		& 450 & 7176.0 & 15.947 & -0.050 & -0.025 \\
Projected network $(P_\beta^\alpha)$	& 450 & 30398.0 & 67.551 & 0.795 & 0.560 \\
\hline
\end{tabular}
\label{tab:Global}
\end{center} 
\end{table}

\begin{table}[t] 
\begin{center}
\scalefont{0.85}
\caption{Structural properties of each layer of the airport transportation multilayer networks.}
\begin{tabular}{|l|c|c|c|c|c|}
\hline
Company & $N$ & $\mathcal{M}$ & $\langle Q^\alpha \rangle$ & $\rho^P(C_\beta^\alpha)$ & $\rho^S(C_\beta^\alpha)$ \\
\hline
Lufthansa	& 106 & 488.0 & 4.604 & -0.668 & -0.473 \\
Ryanair	& 128 & 1202.0 & 9.391 & -0.321 & -0.348 \\
Easyjet	& 99 & 614.0 & 6.202 & -0.428 & -0.470 \\
British Airways	& 65 & 132.0 & 2.031 & -0.775 & -0.754 \\
Turkish Airlines	& 86 & 236.0 & 2.744 & -0.697 & -0.567 \\
Air Berlin	& 75 & 368.0 & 4.907 & -0.501 & -0.434  \\
Air France	& 59 & 138.0 & 2.339 & -0.637 & -0.661 \\
Scandinavian Airlines	& 66 & 220.0 & 3.333 & -0.681 & -0.521 \\
KLM	& 63 & 124.0 & 1.968 & -1.000 & -1.000 \\
Alitalia	& 51 & 186.0 & 3.647 & -0.572 & -0.538 \\
Swiss International Air Lines	& 48 & 120.0 & 2.500 & -0.728 & -0.618 \\
Iberia	& 35 & 70.0 & 2.000 & -0.900 & -0.838 \\
Norwegian Air Shuttle	& 52 & 174.0 & 3.346 & -0.511 & -0.523 \\
Austrian Airlines	& 67 & 144.0 & 2.149 & -0.823 & -0.744 \\
Flybe	& 43 & 198.0 & 4.605 & -0.560 & -0.489 \\
Wizz Air	& 45 & 184.0 & 4.089 & -0.350 & -0.381 \\
TAP Portugal	& 42 & 106.0 & 2.524 & -0.779 & -0.610 \\
Brussels Airlines	& 44 & 86.0 & 1.955 & -1.000 & -1.000 \\
Finnair	& 42 & 84.0 & 2.000 & -0.915 & -0.858 \\
LOT Polish Airlines	& 44 & 110.0 & 2.500 & -0.658 & -0.598 \\
Vueling Airlines	& 36 & 126.0 & 3.500 & -0.438 & -0.456 \\
Air Nostrum	& 48 & 138.0 & 2.875 & -0.571 & -0.569 \\
Air Lingus	& 45 & 116.0 & 2.578 & -0.670 & -0.625 \\
Germanwings	& 44 & 134.0 & 3.045 & -0.628 & -0.482 \\
Panagra Airways	& 45 & 116.0 & 2.578 & -0.625 & -0.593 \\
Netjets	& 94 & 360.0 & 3.830 & -0.106 & 0.107 \\
Transavia Holland	& 40 & 114.0 & 2.850 & -0.585 & -0.535 \\
Niki	& 36 & 74.0 & 2.056 & -0.838 & -0.784 \\
SunExpress	& 38 & 134.0 & 3.526 & -0.797 & -0.542 \\
Aegean Airlines	& 38 & 106.0 & 2.789 & -0.583 & -0.560 \\
Czech Airlines	& 42 & 82.0 & 1.952 & -1.000 & -1.000 \\
European Air Transport	& 53 & 146.0 & 2.755 & -0.416 & -0.423 \\
Malev Hungarian Airlines	& 35 & 68.0 & 1.943 & -1.000 & -1.000 \\
Air Baltic	& 45 & 90.0 & 2.000 & -0.844 & -0.812 \\
Wideroe	& 45 & 180.0 & 4.000 & -0.293 & -0.311 \\
TNT Airways	& 53 & 122.0 & 2.302 & -0.415 & -0.346 \\
Olympic Air 	& 37 & 86.0 & 2.324 & -0.754 & -0.662 \\
\hline
\end{tabular}
\label{tab:Layers}
\end{center} 
\end{table}

\section{Epidemic spreading in correlated multilayer networks}

\begin{figure*}[t]
\subfigure[]{ \includegraphics[width=0.98\columnwidth]{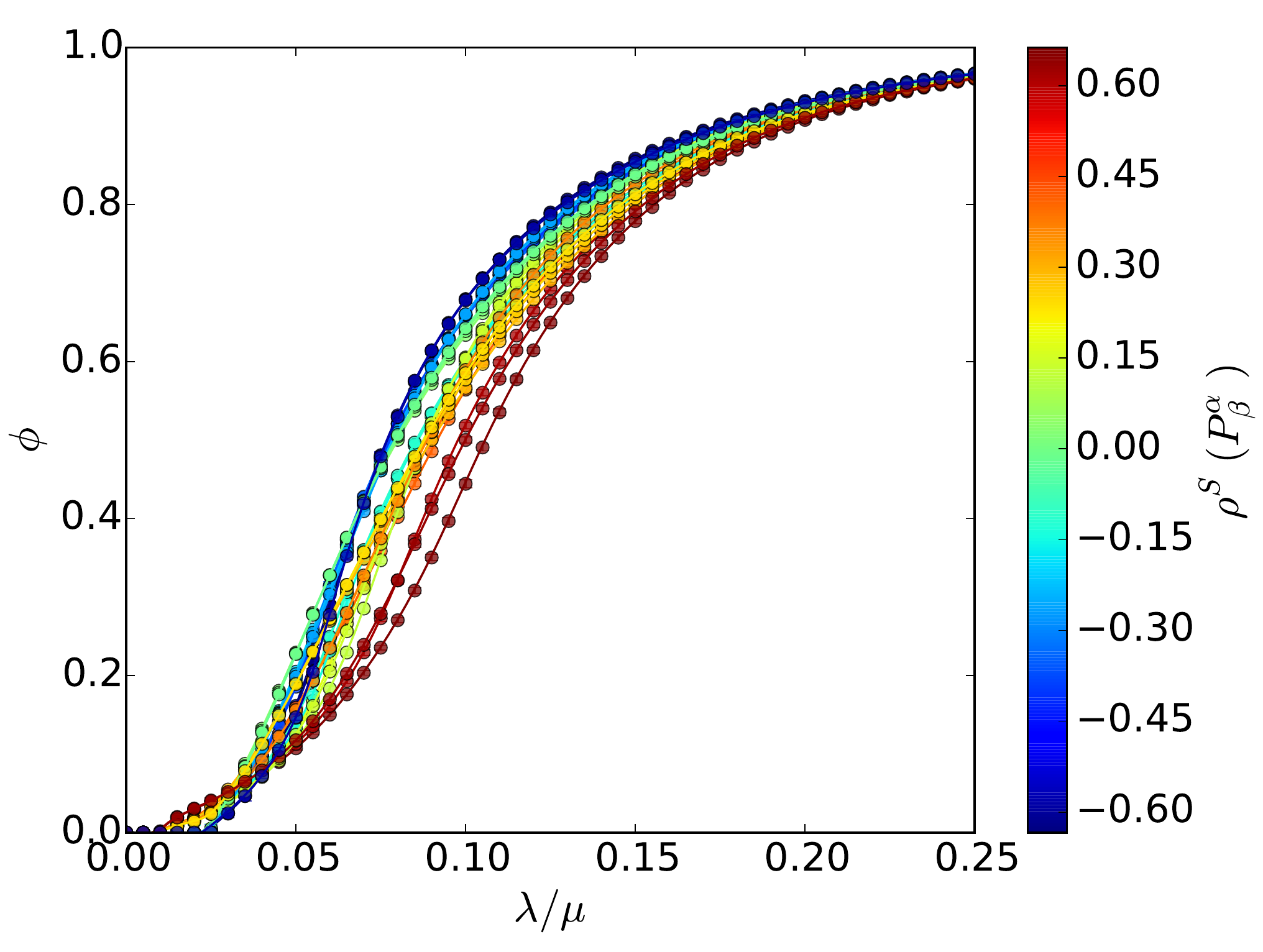} }
\subfigure[]{ \includegraphics[width=0.98\columnwidth]{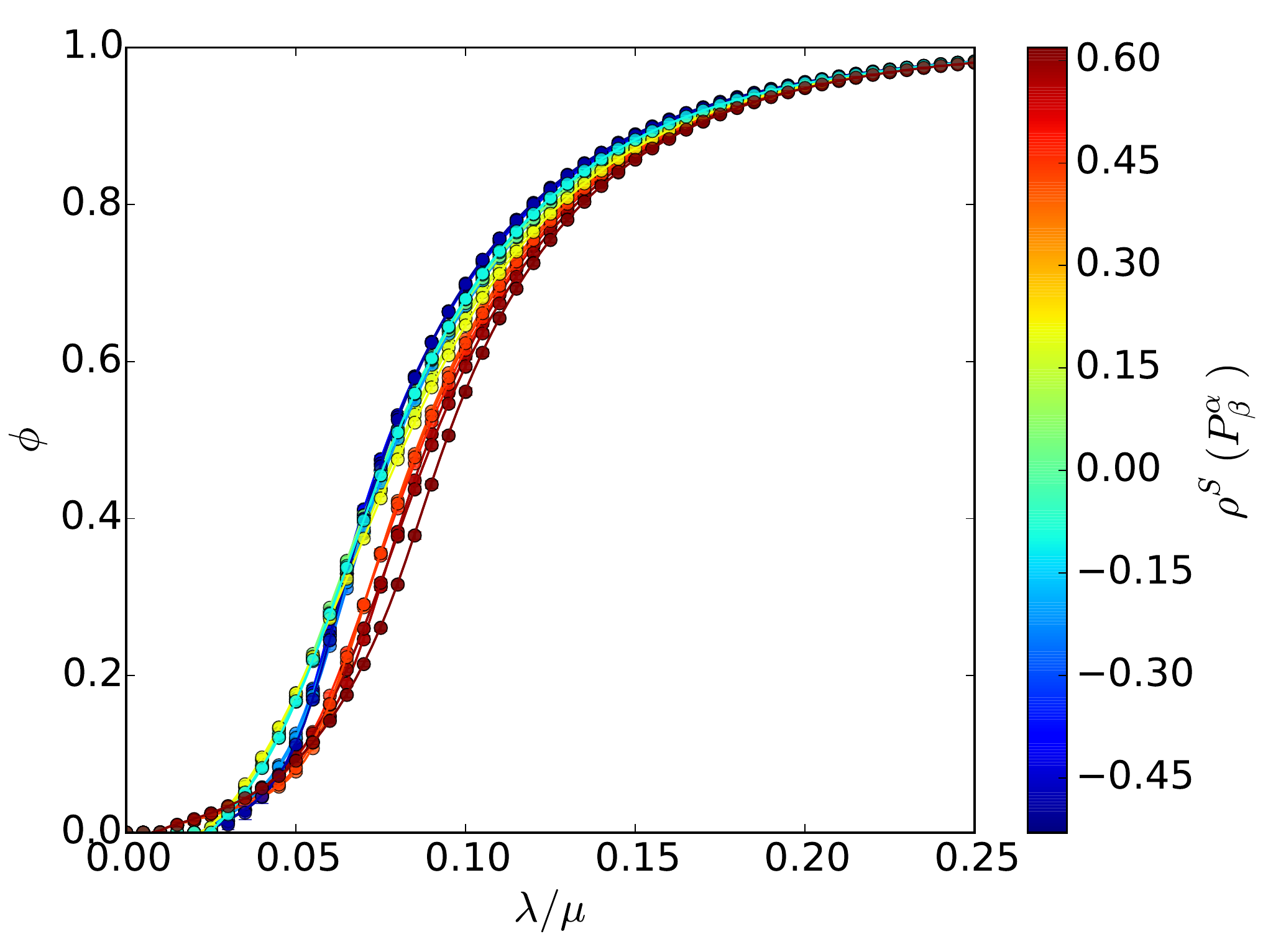} }
\caption{Phase diagram for epidemic spreading on different levels of assortativity. (a) First layer with $m\approx 3$ and $\langle k \rangle \approx 17$, and second with $m \approx 2.8$ and $\langle k \rangle \approx 12$. (b) First layer with $m \approx 3$ and $\langle k \rangle \approx 17$, and second with $m \approx 4.5$ and $\langle k \rangle \approx 13$. All networks are composed by $N = 10^4$ nodes. We adopt $\frac{\gamma}{\lambda} = 2$ and $\mu = 1$. The continuous lines are the analytical solution (Eq.~\ref{Eq.phi}), while the symbols are obtained from Monte Carlo simulations, averaging over $10^2$ runs. The standard deviation is of the size of the symbols.}
\label{Fig:sis}
\end{figure*}

We investigate the effects of degree-degree correlations on epidemic spreading. To this end, we consider a classical SIS (Susceptible-Infected-Susceptible) model, in which nodes can be in one of two states, susceptible or infected~\cite{Satorras14}. Susceptible individuals can be infected if they are in contact with infected individuals, who have already caught the disease and are actively spreading it. Infected individuals get back to the susceptible state with probability $\mu$. Here, we adopt the discrete formulation presented in~\cite{Cozzo:2013}, considering a fully reactive processes (RP), and also perform Monte Carlo simulations of the  epidemic process.

As in \cite{Cozzo:2013}, we consider intra and inter layer spreading. Let's $\lambda$ be the probability of spreading through an intra-layer contact and $\gamma$ the spreading probability through an inter-layer contact. We also assume the possibility of re-infection, that is, an infected individual can be cured and re-infected in the same time interval. Furthermore, it is convenient to consider the ratio between intra-layer and inter-layer spreading probabilities as a constant~\cite{Cozzo:2013}, here we set $\eta = \frac{\gamma}{\lambda} = 2$. However we find similar results considering other ratios. 

To obtain the expression of the macro-state variable in the tensorial notation, we redefine the supra-contact matrix as
\begin{equation} 
\mathcal{R}_{\beta \tilde{\delta}}^{\alpha \tilde{\gamma}}(\lambda, \gamma) = M_{\beta \tilde{\sigma}}^{\alpha \tilde{\eta}} E^{\tilde{\sigma}}_{\tilde{\eta}}(\tilde{\gamma} \tilde{\delta})  \delta^{\tilde{\gamma}}_{\tilde{\delta}} +
\frac{\gamma}{\lambda} M_{\beta \tilde{\sigma}}^{\alpha \tilde{\eta}} E^{\tilde{\sigma}}_{\tilde{\eta}}(\tilde{\gamma} \tilde{\delta}) (U^{\tilde{\gamma}}_{\tilde{\delta}} - \delta^{\tilde{\gamma}}_{\tilde{\delta}}),
\end{equation}
where $E^{\tilde{\sigma}}_{\tilde{\eta}}(\tilde{\gamma} \tilde{\delta}) \in \mathbb{R}^{L \times L}$ indicates the tensor in the canonical basis and $\delta^{\tilde{\gamma}}_{\tilde{\delta}}$ is the Kronecker delta, which is equal to one if $\tilde{\gamma} =\tilde{\delta}$ and zero otherwise.

Denoting the probability of the node $\beta$, on layer $\tilde{\delta}$, becoming infected at time $t$ as $X_{\beta \tilde{\delta}}(t)$, the discrete time evolution equation for this probability is described as
\begin{eqnarray} \label{eq:sis_disc}
 X_{\beta \tilde{\delta}}(t+1)&=&(1 - X_{\beta \tilde{\delta}}(t))(1 - q_{\beta \tilde{\delta}}(t))+(1 - \mu )X_{\beta \tilde{\delta}}(t)\nonumber\\
&+&\mu(1 - q_{\beta \tilde{\delta}}(t))X_{\beta \tilde{\delta}}(t),
\end{eqnarray}
where the probability that a node will not be infected by any of its neighbors at time $t$ is given as
\begin{equation}
 q_{\beta \tilde{\delta}}(t) = \prod_{\alpha} \prod_{\tilde{\gamma}} \left( 1 - \lambda \mathcal{R}^{\beta \tilde{\delta}}_{\alpha \tilde{\gamma}}(\lambda, \gamma) X_{\alpha \tilde{\gamma}} \right).
\end{equation}
Observe that in Eq.~\ref{eq:sis_disc}, the indices $\beta \tilde{\delta}$ are not dummy and there is no summation on it. A more formal notation would be obtained substituting $X_{\beta \tilde{\delta}}$ by $X_{\eta \tilde{\sigma}} E^{\eta \tilde{\sigma}}(\beta \tilde{\delta})$. The implicit summation has only one term different from zero, which is $X_{\beta \tilde{\delta}}$.

Finally, the macro-state variable is given as
\begin{equation}\label{Eq.phi}
 \phi = \frac{1}{LN} X_{\beta \tilde{\delta}} U^{\beta \tilde{\delta}},
\end{equation}
where $U^{\beta \tilde{\delta}} \in \mathbb{R}^{N \times L}$ is the all one tensor. In other words it is an average over all the individuals. Observe that our equations are exactly the same presented in~\cite{Cozzo:2013}, but here we consider the tensorial notation.

In addition to the analytical approach we also perform Monte Carlo simulations to evaluate the influence of degree-degree correlations on epidemic spreading. The simulations are performed in a synchronous manner, i.e., every node changes its state at the same time and the events between $t$ and $t+1$ are assumed to occur at the same time. In this way, at each time step, every spreader is cured with probability $\mu$. After that, every infected individual contacts all its neighbors, thus representing a fully reactive processes (RP). However, note that a cured spreader can still propagate the disease, because its state changes only at the end of the time step. This procedure enables the occurrence of reinfections. After the contact, the disease spreading can occur in two different ways: (i) for inter-layers, where the spreading takes place with probability $\lambda$, or (ii) for intra-layers, where the spreading occurs with probability $\gamma$. 

In order to quantify the effect of degree-degree correlations on the spreading process, we generate two scale-free networks, with degree distribution $P(k) \approx k^{-m}$, according to the configuration model~\cite{Viger2005}. The first layer has $m \approx 3$ and $\langle k \rangle \approx 17$, whereas the other one we evaluate in two different configurations: (i) $m \approx 4.5$ and $\langle k \rangle \approx 13$ and (ii)  $m \approx 2.8$ and $\langle k \rangle \approx 12$. Both networks are composed by $N = 10^4$ nodes. 

On the other hand, to control the level of degree-degree correlations in random networks, we consider a simulated annealing algorithm~\cite{Kirkpatrick1983}. This algorithm is based on two functions, i.e., (i) the perturbation function, which changes the system configuration, and (ii) the energy function, which is minimized. In our case, the perturbation function is a rewiring procedure that preserves the degree distribution of the network, but changes the large-scale degree-degree correlations. The energy function is defined as $E_t = c(\rho_t + 1)$, where $\rho_t$ is the network assortativity at time $t$ and $c$ is a constant related to the level of degree-degree correlation, i.e., $c = -1$ if the goal is to obtain an assortative network or $c = 1$ if the goal is a disassortative network.

Given an initial network configuration, an initial temperature, $T$ and a cooling factor $\alpha$, the algorithm can be described by the following steps: (i) the energy function is initialized as $E_0$;  (ii) while the number of iterations are less than a threshold or the optimal solution is not found (or good solution, given a tolerance) the following steps are performed: (iii) a rewiring preserving the degree distribution is executed, according to our perturbation function; (iv) the new energy function, $E_{t+1}$, is calculated; (v) if $E_t - E_{t+1} < 0$ or $\exp \left( \frac{-(E_t - E_{t+1})}{T} \right) < U(0,1)$ , where $U(0,1)$ is a random number sampled from a uniform distribution in $[0,1]$, then the new solution is accepted; (vi) the temperature is updated, $T = \alpha T$; and (vii) increment the iteration counter. Observe that a worse state than the current one can be accepted with a probability $\exp \left( \frac{-(E_t - E_{t+1})}{T} \right)$. This mechanism allows the system to avoid local minima. Following this procedure, we can generate random networks with a defined level of degree-degree correlation.

Thus, using the simulated annealing algorithm above, we tune the assortativity on the individual layers. We can have three different configurations for each layer, i.e.,  (i) one assortative, (ii) one disassortative and (iii) one non-assortative. Those individual layers are connected, forming a multiplex network. In this case, we can have three different configurations: (i) assortative: densely connected nodes from one layer is connected to densely connected nodes in the other layer, (ii) disassortative: hubs in one layer are connected to low degree nodes in the other layer, and (iii) random: nodes in different layers are randomly connected. In this way, our data set is composed by 27 multiplex networks presenting different levels of assortativity. We also consider $\eta = \frac{\gamma}{\lambda} = 2$ and $\mu = 1$. Similar results are found for different values of $\eta$. 

Figure~\ref{Fig:sis} shows the simulations of the SIS dynamics on top of multiplex networks with different levels of assortativity. We can see a good agreement between the Monte Carlo simulation and the theoretical macro-state variable (see Eq.~\ref{Eq.phi}), although we assume that there is no correlation among the state of each random variable. Each network has different values of the epidemic threshold and also exhibits different behaviors near the threshold. Indeed, the epidemic threshold for assortative networks is at a lower transmission probability. This happens because the disease has a faster initial growth rate and a shorter duration in assortative networks than in disassortative networks. Nevertheless, disassortative networks show higher values of $\phi$ for larger values of $\gamma / \lambda$. This result agrees with the analysis of single layer networks in~\cite{Kiss08}. Notice that the same behavior is observed for Figures~\ref{Fig:sis}(a) and~\ref{Fig:sis}(b), although in (a) the network is more heterogeneous than in (b). In fact, for more heterogeneous networks, the influence of degree-degree correlations is reduced.

\section{Conclusions}

In this paper we have generalized the metric used to calculate assortativity of multilayer networks. Our approach consists of reducing the dimension of the adjacency tensor and applying the Pearson correlation coefficient on the extremes of each edge. We follow the tensorial approach, which help us to have a compact, algorithmic and general formulation, covering various topological representations and possibilities, such as overlay and projected networks, and also pair-wise measurements. The calculation of the Spearman rank correlation is also possible from our formulation.

In the study of the airport transportation network, we verified that the individual analysis of the overlay or the projected networks can yield misleading conclusions. Indeed, as shown in Table~\ref{tab:Global} the assortativity values for the projected and overlay networks are different. The overlay network shows a small disassortative behavior, while the projected graph is highly assortative. This indicates that the main contribution to the assortativity of the projected network is given by self-edges, i.e., hub airports that are present in many different layers. On the other hand, individually, each layer is disassortative, with a value much higher in module than the one obtained for the overlay network. This is because companies tend to have one big hub from which connections to many other airports are established, but, at the same time, they tend to have direct flights to the hubs of other companies. This interpretation of the data is also confirmed in Figure~\ref{Fig:Airport_projection_rho}, in which we observe a high negative value along the diagonal of the matrix, i.e., a strong disassortative behavior of isolated layers. In addition, we can see relative smaller negative values for the elements out of the diagonal, i.e., a relative weaker disassortative behavior, representing pair-wise correlations between different layers, i.e. companies. Furthermore, the comparison of the \textit{P-assortativity} and the \textit{O-assortativity} matrices also emphasizes the importance of the two different analysis, where the first is slightly higher due to the self-edges. Finally, the network of layers shows an assortative behavior, suggesting again that the main airline companies share similar airports. 

Finally, we studied the effects of degree-degree correlations on epidemic spreading. The results obtained from Monte Carlo simulations and theoretical analysis using a Markov Chain formulation in terms of tensors show that the level of assortativity and heterogeneity between layers influence the spreading process. More specifically, we verified that assortative networks show a smaller epidemic threshold, and that the disease has a faster initial growth rate in these networks, but a shorter duration. On the contrary, the fraction of infected individuals is larger in disassortative networks. Finally, we have also shown that degree-degree correlations have a larger impact on the spreading dynamics when the coupled networks have similar levels of heterogeneity.

\textbf{Acknowledgment}
FAR acknowledge CNPq (grant 307974/2013-8) and Fapesp (grant 2013/26416-9) for financial support. GFA acknowledges Fapesp for the sponsorship provided (grants 2012/25219-2 and 2015/07463-1). GFA acknowledges Felipe Montefuscolo for fruitful comments. YM is partially supported by the EC FET-Proactive Project PLEXMATH (grant 317614).

\bibliographystyle{apsrev}
\bibliography{paper}

\end{document}